\documentclass[a4paper]{article}

\usepackage{fixltx2e}
\usepackage{isomath}

\usepackage[utf8]{inputenc}
\usepackage[T1]{fontenc}
\usepackage[british]{babel}

\makeatletter
\useshorthands{"}%
\defineshorthand{"-}{\nobreak-\bbl@allowhyphens}
\defineshorthand{"=}{\nobreak--\bbl@allowhyphens}
\makeatother

\usepackage{amsmath}
\usepackage{amsfonts}
\usepackage{natbib}
\usepackage{proof}
\usepackage{tikz}
\usetikzlibrary{calc}
\usepackage{url}

\usepackage{microtype}

\usepackage{hyperref}

\newcommand*{\INFER}[3][]{\ensuremath{\infer[\text{\footnotesize #1}]{\mathstrut #2}{\mathstrut #3}}}

\newcommand*{\GRAPHR}[2]{%
	\begin{tikzpicture}[xscale=0.5, yscale=0.5]
		\path[draw, white] (0, 0) rectangle (1, 1);
		\path[thick, ->] (0, 0) edge[out=90, in=90, looseness=3] (1, 0);
		\node[anchor=north] at (0, 0) {$\mathstrut #1$};
		\node[anchor=north] at (1, 0) {$\mathstrut #2$};
	\end{tikzpicture}%
}

\newcommand*{\GRAPHL}[2]{%
	\begin{tikzpicture}[xscale=0.5, yscale=0.5]
		\path[draw, white] (0, 0) rectangle (1, 1);
		\path[thick, <-] (0, 0) edge[out=90, in=90, looseness=3] (1, 0);
		\node[anchor=north] at (0, 0) {$\mathstrut #1$};
		\node[anchor=north] at (1, 0) {$\mathstrut #2$};
	\end{tikzpicture}%
}

\newcommand*{\GRAPHH}[2]{%
	\begin{tikzpicture}[xscale=0.5, yscale=0.5]
		\path[draw, white] (0, 0) rectangle (1, 1);
		\draw[thick, help lines] (0, 0) rectangle (1, 0.75);
		\node[anchor=north] at (0, 0) {$\mathstrut #1$};
		\node[anchor=north] at (1, 0) {$\mathstrut #2$};
	\end{tikzpicture}%
}

\newcommand*{\SEQR}[2]{%
	\begin{tikzpicture}[xscale=0.5, yscale=0.5]
		\path[draw, white] (0, 0) rectangle (3.5, 1);
		\path[thick, ->] (0, 0) edge[out=90, in=90, looseness=3] (1, 0);
		\path[thick, ->] (2.5, 0) edge[out=90, in=90, looseness=3] (3.5, 0);
		\node at (1.75, 0.5) {$\mathstrut\cdots$};
		\node[anchor=north] at (0, 0) {$\mathstrut #1$};
		\node[anchor=north] at (3.5, 0) {$\mathstrut #2$};
	\end{tikzpicture}%
}

\newcommand*{\SEQL}[2]{%
	\begin{tikzpicture}[xscale=0.5, yscale=0.5]
		\path[draw, white] (0, 0) rectangle (3.5, 1);
		\path[thick, <-] (0, 0) edge[out=90, in=90, looseness=3] (1, 0);
		\path[thick, <-] (2.5, 0) edge[out=90, in=90, looseness=3] (3.5, 0);
		\node at (1.75, 0.5) {$\mathstrut\cdots$};
		\node[anchor=north] at (0, 0) {$\mathstrut #1$};
		\node[anchor=north] at (3.5, 0) {$\mathstrut #2$};
	\end{tikzpicture}%
}

\newcommand*{\SEQM}[2]{%
	\begin{tikzpicture}[xscale=0.5, yscale=0.5]
		\path[draw, white] (0, 0) rectangle (3.5, 1);
		\path[thick] (0, 0) edge[out=90, in=90, looseness=3] (1, 0);
		\path[thick] (2.5, 0) edge[out=90, in=90, looseness=3] (3.5, 0);
		\node at (1.75, 0.5) {$\mathstrut\cdots$};
		\node[anchor=north] at (0, 0) {$\mathstrut #1$};
		\node[anchor=north] at (3.5, 0) {$\mathstrut #2$};
	\end{tikzpicture}%
}

\newcommand*{\SEQU}[2]{%
	\begin{tikzpicture}[xscale=0.5, yscale=0.5]
		\path[draw, white] (0, 0) rectangle (3.5, 1);
		\path[thick] (0, 0) edge[out=90, in=90, looseness=3] (1, 0);
		\path[thick] (2.5, 0) edge[out=90, in=90, looseness=3] (3.5, 0);
		\node at (1.75, 0.5) {$\mathstrut\cdots$};
		\node at (1.75, 0.5) {$//$};
		\node[anchor=north] at (0, 0) {$\mathstrut #1$};
		\node[anchor=north] at (3.5, 0) {$\mathstrut #2$};
	\end{tikzpicture}%
}

\begin{document}

\title{Tabulation of Noncrossing Acyclic Digraphs}

\author{%
	Marco Kuhlmann\\
	Department of Computer and Information Science\\
	Linköping University, Sweden}

\maketitle

\section{Introduction}

\begin{figure}
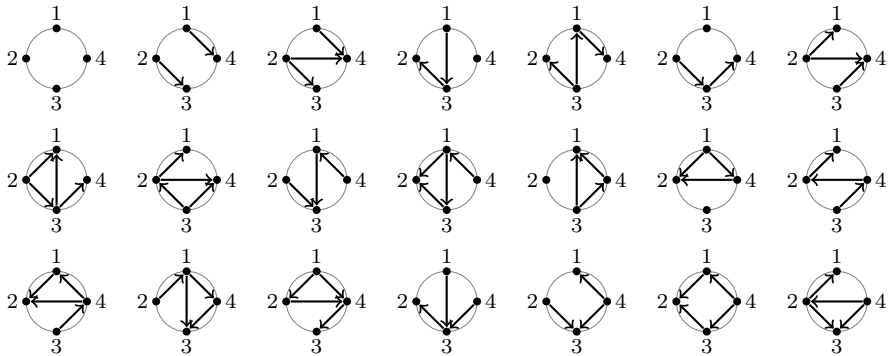

	\centering\footnotesize
	\input ncdags-examples
	\caption{Some (out of 335) noncrossing acyclic digraphs of size~4.}
	\label{fig:ncdags-examples}
\end{figure}

A \emph{noncrossing graph} is a graph with its vertices drawn on a
circle and its edges drawn in the interior such that no two edges
cross each other.
This note is concerned with \emph{noncrossing acyclic digraphs}.
Examples of such structures are given in
Figure~\ref{fig:ncdags-examples}.
I present an algorithm that, given a number $n \geq 1$, computes a
compact representation of the set of all noncrossing acyclic digraphs
with $n$ nodes.
This compact representation can be used as the basis for a wide range
of dynamic programming algorithms on these graphs.
As an illustration, along with this note I am releasing the
implementation of an algorithm for counting the number of noncrossing
acyclic digraphs of a given
size.\footnote{\url{https://github.com/khlmnn/ncdags}} This number is
given by the following integer sequence (starting with $n = 1$); this
is A246756 in the OEIS \citep{oeis}:
\begin{center}
  1, 3, 25, 335, 5521, 101551, 1998753, 41188543, 877423873, 19166868607, \dots
\end{center}
Another application of the tabulation technique is in semantic
dependency parsing \citep{oepen2014broad}, where it can be used to
compute the highest"-scoring dependency graph for a given sentence
under an edge"-factored model.

\section{Preliminaries}

Before presenting the tabulation I introduce some terminology for
special types of noncrossing acylic digraphs, as well as a set of
operations on graphs that will be useful for the understanding of the
technique.

\subsection{Classification}
\label{sec:Classification}

The proposed tabulation is based on a classification of noncrossing
acyclic digraphs into 7 different types.
To simplify the presentation I only consider graphs with at least $2$
nodes.

A graph is called \emph{edge"-covered} if there is an edge connecting
its extremal vertices.
Thus there are 2~types of edge"-covered graphs:
\begin{itemize}
\item The covering edge goes from the minimal vertex to the maximal
  vertex.
  In this case, I say that the graph is \emph{minmax"-covered}.
\item The covering edge goes from the minimal vertex to the maximal
  vertex.
  In this case, I say that the graph is \emph{maxmin"-covered}.
\end{itemize}
If a graph is not edge"-covered, I distinguish two cases depending on
whether or not the graph is weakly connected -- that is, whether there
exists a path (consisting of two or more edges) between the extremal
vertices.
In the following I use the term \emph{connected} in the sense `weakly
connected but not edge"-covered'.
I distinguish 3~types of connected graphs:
\begin{itemize}
\item There is a directed path from the minimal vertex to the maximal
  vertex.
  In this case, I say that the graph is \emph{minmax"-connected}.
\item There is a directed path from the maximal vertex to the minimal
  vertex.
  In this case, I say that the graph is \emph{maxmin"-connected}.
  Note that because of acyclicity, a graph cannot be both
  minmax"-connected and maxmin"-connected.
\item There is no directed path between the two extremal vertices,
  implying that there is a path consisting of edges with mixed
  directions.
  In this case, I say that the graph is \emph{mix"-connected}.
\end{itemize}
The last two types are the graphs that are neither edge"-covered nor
connected.
In the following I refer to these graphs as \emph{unconnected}.
I distinguish 2~types:
\begin{itemize}
\item The graph has $2$ nodes.
  In this case, the graph is uniquely determined; it is the graph with
  $2$ nodes and no edges.
  I refer to this graph as the \emph{elementary graph}.
\item The graph has more than $2$ nodes.
  I say that the graph is \emph{non"-elementary}.
\end{itemize}
Note that this classification is exhaustive, meaning that every
noncrossing acyclic digraph falls into (exactly) one of the 7 classes.

\subsection{Operations}
\label{sec:Operations}

The proposed tabulation takes an algebraic view on noncrossing acyclic
digraphs where every graph is composed from `smaller' graphs by means
of three operations:
\begin{itemize}
\item \emph{Concatenate} two graphs, identifying the last vertex of
  the first graph with the first vertex of the second graph.
  This operation is perhaps easiest illustrated by drawing the graphs
  on a straight line rather than on a circle.
  (With this layout, the non"-crossing condition means that the edges
  can be drawn in the half"-plane above the line without crossings.)
  \begin{displaymath}
    \text{concatenating}
    \quad
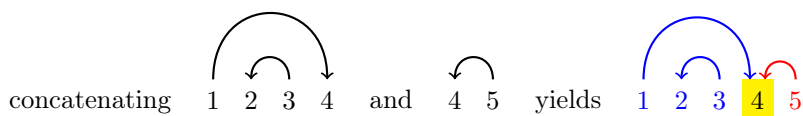
    \begin{tikzpicture}[xscale=0.5, baseline=(n1.base)]
      \node (n1) at (0, 0) {$\mathstrut 1$};
      \node (n2) at (1, 0) {$\mathstrut 2$};
      \node (n3) at (2, 0) {$\mathstrut 3$};
      \node (n4) at (3, 0) {$\mathstrut 4$};
      \path[thick, ->] (n1) edge[out=90, in=90] (n4);
      \path[thick, ->] (n3) edge[out=90, in=90] (n2);
    \end{tikzpicture}
    \quad
    \text{and}
    \quad
    \begin{tikzpicture}[xscale=0.5, baseline=(n1.base)]
      \node (n4) at (0, 0) {$\mathstrut 4$};
      \node (n5) at (1, 0) {$\mathstrut 5$};
      \path[thick, ->] (n5) edge[out=90, in=90] (n4);
    \end{tikzpicture}
    \quad
    \text{yields}
    \quad
    \begin{tikzpicture}[xscale=0.5, baseline=(n1.base)]
      \node[blue] (n1) at (0, 0) {$\mathstrut 1$};
      \node[blue] (n2) at (1, 0) {$\mathstrut 2$};
      \node[blue] (n3) at (2, 0) {$\mathstrut 3$};
      \node[fill=yellow] (n4) at (3, 0) {$\mathstrut 4$};
      \node[red] (n5) at (4, 0) {$\mathstrut 5$};
      \path[thick, ->, blue] (n1) edge[out=90, in=90] ($(n4.north) - (0.2, 0)$);
      \path[thick, ->, blue] (n3) edge[out=90, in=90] (n2);
      \path[thick, ->, red] (n5) edge[out=90, in=90] ($(n4.north) + (0.2, 0)$);
    \end{tikzpicture}
  \end{displaymath}
  Here the vertices and edges contributed by the first graph are drawn
  in blue, those contributed by the second graph are drawn in red, and
  the joint vertex (simultaneously the last vertex of the first graph
  and the first vertex of the second graph) is highlighted in yellow.

\item \emph{Cover} a graph by adding a new edge (with two possible
  directions) between the first vertex and the last vertex.
  In the following illustration, the new edges are drawn in red:
  \begin{displaymath}
    \text{covering}
    \quad
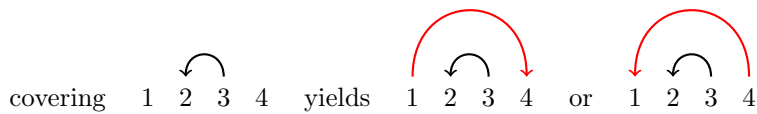
    \begin{tikzpicture}[xscale=0.5, baseline=(n1.base)]
      \node (n1) at (0, 0) {$\mathstrut 1$};
      \node (n2) at (1, 0) {$\mathstrut 2$};
      \node (n3) at (2, 0) {$\mathstrut 3$};
      \node (n4) at (3, 0) {$\mathstrut 4$};
      \path[thick, ->] (n3) edge[out=90, in=90] (n2);
    \end{tikzpicture}
    \quad
    \text{yields}
    \quad
    \begin{tikzpicture}[xscale=0.5, baseline=(n1.base)]
      \node (n1) at (0, 0) {$\mathstrut 1$};
      \node (n2) at (1, 0) {$\mathstrut 2$};
      \node (n3) at (2, 0) {$\mathstrut 3$};
      \node (n4) at (3, 0) {$\mathstrut 4$};
      \path[thick, ->, red] (n1) edge[out=90, in=90] (n4);
      \path[thick, ->] (n3) edge[out=90, in=90] (n2);
    \end{tikzpicture}
    \quad
    \text{or}
    \quad
    \begin{tikzpicture}[xscale=0.5, baseline=(n1.base)]
      \node (n1) at (0, 0) {$\mathstrut 1$};
      \node (n2) at (1, 0) {$\mathstrut 2$};
      \node (n3) at (2, 0) {$\mathstrut 3$};
      \node (n4) at (3, 0) {$\mathstrut 4$};
      \path[thick, <-, red] (n1) edge[out=90, in=90] (n4);
      \path[thick, ->] (n3) edge[out=90, in=90] (n2);
    \end{tikzpicture}
  \end{displaymath}
\end{itemize}
Note that the set of noncrossing acyclic digraphs is not closed under
these operations.
In particular, the cover operations may introduce cycles and even
multiple edges.

\section{Tabulation}

I present the proposed tabulation as a deduction system in the sense
of \citet{shieber1995principles}.
Tabulation is viewed as a deductive process in which rules of
inference are used to derive statements about sets of graphs from
other such statements.
Statements are represented by formulas called \emph{items}.

\paragraph{Notation}

Recall that I assume that $n \geq 2$.
In the following, $1 \leq i \leq j \leq k \leq n$.

\subsection{Items}

Following the classification given in
Section~\ref{sec:Classification}, the items of the deduction system
take one of 7 possible forms.
I represent these items using a graphical notation that is intended to
be mnemonic.
\begin{itemize}
\item \textit{Items for edge"-covered graphs.} For $j - i \geq 1$:
  \begin{displaymath}
    \GRAPHR{i}{j}
    \qquad
    \GRAPHL{i}{j}
  \end{displaymath}
  The intended interpretation of these items is: `It is possible to
  construct an edge"-covered noncrossing acyclic digraph on the
  vertices $i, \dots, j$.'
  
\item \textit{Items for connected graphs.} For $j - i \geq 2$:
  \begin{displaymath}
    \SEQR{i}{j}
    \qquad
    \SEQL{i}{j}
    \qquad
    \SEQM{i}{j}
  \end{displaymath}
  The intended interpretation of these items is: `It is possible to
  construct an minmax"-connected, maxmin"-connected, mix"-connected
  noncrossing acyclic digraph on the vertices $i, \dots, j$.'
  
\item \textit{Items for elementary graphs.} For $j - i = 1$:
  \begin{displaymath}
    \GRAPHH{i}{j}
  \end{displaymath}
  The intended interpretation of these items is: `The elementary graph on the vertices $i, j$ is a noncrossing acyclic digraph.'
  
\item \textit{Items for unconnected graphs.} For $j - i \geq 2$:
  \begin{displaymath}
    \SEQU{i}{j}
  \end{displaymath}
  The intended interpretation of these items is: `It is possible to
  construct an unconnected noncrossing acyclic digraph on the vertices
  $i, \dots, j$.'
\end{itemize}

\subsection{Axioms}

The axioms of the deduction system are the items for the elementary
graphs.

\subsection{Rules}

The deduction system has 26 rules.
Each of these rules simulates a concatenation or cover operation on
the 7 different types of noncrossing acyclic digraphs specified in
Section~\ref{sec:Classification}.

\paragraph{Concatenate two edge"-covered graphs}

The first four rules simulate the concatenation of two edge"-covered
graphs.
The result of such a concatenation is a connected graph:
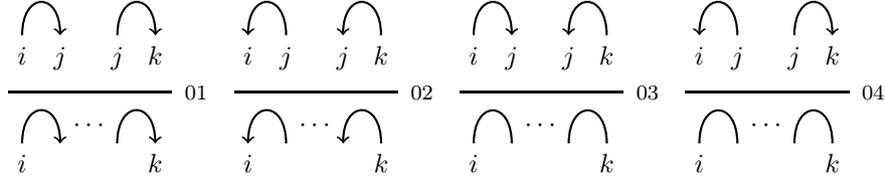
\begin{displaymath}
  \INFER[01]{\SEQR{i}{k}}{\GRAPHR{i}{j} & \GRAPHR{j}{k}}
  \quad
  \INFER[02]{\SEQL{i}{k}}{\GRAPHL{i}{j} & \GRAPHL{j}{k}}
  \quad
  \INFER[03]{\SEQM{i}{k}}{\GRAPHR{i}{j} & \GRAPHL{j}{k}}
  \quad
  \INFER[04]{\SEQM{i}{k}}{\GRAPHL{i}{j} & \GRAPHR{j}{k}}
\end{displaymath}
For instance, rule~03 states that the concatenation of a
minmax"-covered graph on the vertices $i, \dots, j$ and an
maxmin"-covered graph on the vertices $j, \dots, k$ yields a
mix"-connected graph on the vertices $i, \dots, k$.

\paragraph{Concatenate an edge"-covered graph and the elementary graph}

The next rules simulate the concatenation of an edge"-covered graph
and the elementary graph.
The result of such a concatenation is an unconnected graph.
There are 4~cases:
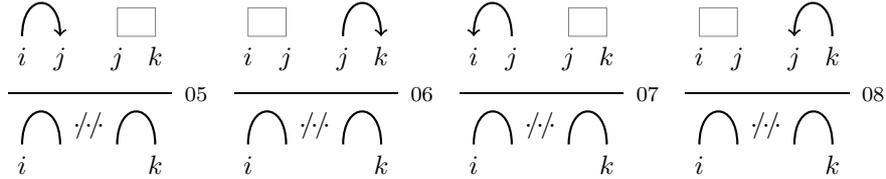
\begin{displaymath}
  \INFER[05]{\SEQU{i}{k}}{\GRAPHR{i}{j} & \GRAPHH{j}{k}}
  \quad
  \INFER[06]{\SEQU{i}{k}}{\GRAPHH{i}{j} & \GRAPHR{j}{k}}
  \quad
  \INFER[07]{\SEQU{i}{k}}{\GRAPHL{i}{j} & \GRAPHH{j}{k}}
  \quad
  \INFER[08]{\SEQU{i}{k}}{\GRAPHH{i}{j} & \GRAPHL{j}{k}}
\end{displaymath}

\paragraph{Concatenate a connected graph and an edge"-covered graph}

The following rules simulate the concatenation of a connected graph
and an edge"-covered graph.
The result of such a concatenation is a connected graph.
There are 6 cases; I group them based on the type of the first
argument of the concatenation operation.

\begin{trivlist}
\item\relax \textit{Group~1: The first argument is minmax"-connected}
  \begin{displaymath}
    \INFER[09]{\SEQR{i}{k}}{\SEQR{i}{j} & \GRAPHR{j}{k}}
    \qquad
    \INFER[10]{\SEQM{i}{k}}{\SEQR{i}{j} & \GRAPHL{j}{k}}
  \end{displaymath}
  
\item\relax \textit{Group~2: The first argument is maxmin"-connected}
  \begin{displaymath}
    \INFER[11]{\SEQM{i}{k}}{\SEQL{i}{j} & \GRAPHR{j}{k}}
    \qquad
    \INFER[12]{\SEQL{i}{k}}{\SEQL{i}{j} & \GRAPHL{j}{k}}
  \end{displaymath}
  
\item\relax \textit{Group~3: The first argument is mix"-connected}
  \begin{displaymath}
    \INFER[13]{\SEQM{i}{k}}{\SEQM{i}{j} & \GRAPHR{j}{k}}
    \qquad
    \INFER[14]{\SEQM{i}{k}}{\SEQM{i}{j} & \GRAPHL{j}{k}}
  \end{displaymath}
\end{trivlist}

\paragraph{Concatenate a connected graph and the elementary graph}

The next rules simulate the concatenation of a connected graph and the
elementary graph.
The result of such a concatenation is an unconnected graph.
There are 3~cases:
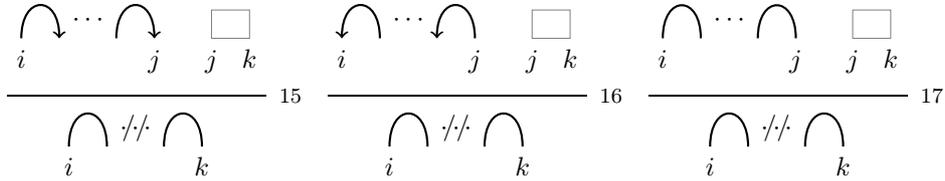
\begin{displaymath}
  \INFER[15]{\SEQU{i}{k}}{\SEQR{i}{j} & \GRAPHH{j}{k}}
  \quad
  \INFER[16]{\SEQU{i}{k}}{\SEQL{i}{j} & \GRAPHH{j}{k}}
  \quad
  \INFER[17]{\SEQU{i}{k}}{\SEQM{i}{j} & \GRAPHH{j}{k}}
\end{displaymath}

\paragraph{Concatenate to an unconnected graph}

The next rules simulate the concatenation to an unconnected graph.
The result of such a concatenation is another unconnected graph.
I consider 3~cases:
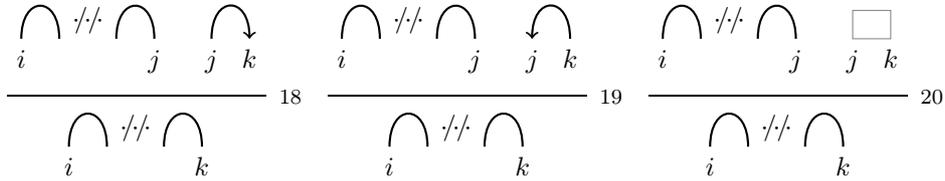
\begin{displaymath}
  \INFER[18]{\SEQU{i}{k}}{\SEQU{i}{j} & \GRAPHR{j}{k}}
  \quad
  \INFER[19]{\SEQU{i}{k}}{\SEQU{i}{j} & \GRAPHL{j}{k}}
  \quad
  \INFER[20]{\SEQU{i}{k}}{\SEQU{i}{j} & \GRAPHH{j}{k}}
\end{displaymath}

\paragraph{Cover a graph}

The rules in the final set simulate the cover operations.
The result of such an operation is an edge"-covered graph.
There are 6~cases; I group them based on the direction of the covering
edge.
\begin{trivlist}
\item\relax \textit{Group~1: The covering edge goes from the minimal vertex to the maximal vertex}
  \begin{displaymath}
    \INFER[21]{\GRAPHR{i}{j}}{\SEQR{i}{j}}
    \qquad
    \INFER[22]{\GRAPHR{i}{j}}{\SEQM{i}{j}}
    \qquad
    \INFER[23]{\GRAPHR{i}{j}}{\SEQU{i}{j}}
  \end{displaymath}
  
\item\relax \textit{Group~2: The covering edge goes from the maximal vertex to the minimal vertex}
  \begin{displaymath}
    \INFER[24]{\GRAPHL{i}{j}}{\SEQL{i}{j}}
    \qquad
    \INFER[25]{\GRAPHL{i}{j}}{\SEQM{i}{j}}
    \qquad
    \INFER[26]{\GRAPHL{i}{j}}{\SEQU{i}{j}}
  \end{displaymath}
\end{trivlist}

This completes the presentation of the rules.

\subsection{Goal Items}

In contrast to the deduction systems of \citet{shieber1995principles},
the proposed tabulation does not have a unique goal item but
7~different goal items, corresponding to the 7~types of noncrossing
acyclic digraphs (Section~\ref{sec:Classification}).
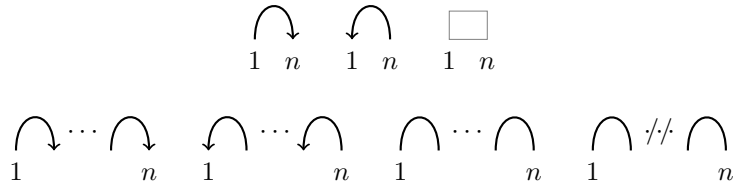
\begin{gather*}
  \GRAPHR{1}{n}
  \quad
  \GRAPHL{1}{n}
  \quad
  \GRAPHH{1}{n}
  \\
  \SEQR{1}{n}
  \quad
  \SEQL{1}{n}
  \quad
  \SEQM{1}{n}
  \quad
  \SEQU{1}{n}
\end{gather*}

\subsection{Properties}

While I shall not provide a complete formal analysis of the
tabulation, I briefly mention some crucial properties:
\begin{itemize}
\item The runtime of the tabulation is in $O(n^3)$ and the space
  required for it is in $O(n^2)$.
  This can be seen by counting the number of possible instances of
  each inference rule and item.
	
\item The deduction system is \emph{sound}, meaning that for each
  rule, if the statements encoded by the antecedents hold, then the
  statement encoded by the consequent holds as well.
  To see this, one can check the soundness of each rule.
  
\item The deduction system is \emph{complete}, meaning that every
  noncrossing acyclic digraph can be constructed in a way that can be
  simulated by the inference rules.
  The completeness argument starts from the observation that the
  classification given in Section~\ref{sec:Classification} is
  exhaustive, and then checks for each rule that undoing the operation
  simulated by that rule decomposes a graph represented by the
  consequent item into the graphs represented by the antecedents.
  
\item Every noncrossing acyclic digraph has a \emph{unique} derivation
  in the deduction system.
  This property is useful because it means that we do not need to
  distinguish between graphs and their derivations.
  In particular, we can count graphs by counting their derivations.
  The uniqueness argument makes use of the observations that the graph
  types distinguished in Section~\ref{sec:Classification} are
  non"-overlapping, and that the backward application of rules is
  deterministic in the sense that for each graph there is at most one
  rule in which this graph appears as the consequent item.
\end{itemize}

\section{Derived Tabulations}

I conclude this note by noting how tabulation techniques for other
classes of noncrossing graphs can be derived from the proposed
technique.

\subsection{Enforcing Weak Connectivity}

To obtain a deduction system for \emph{weakly connected} noncrossing
acyclic digraphs one removes rule 20 (concatenate an unconnected graph
and the elementary graph) and deletes the item for unconnected graphs
from the list of goal items.
Along with this change comes a revised intended interpretation for the
items for unconnected graphs: `It is possible to construct a
noncrossing acyclic digraph on the vertices $i, \dots, j$ that is not
edge"-covered and has exactly two weakly connected components.'
The integer sequence for weakly connected noncrossing acyclic digraphs
is:
\begin{flushleft}
  1, 2, 18, 242, 3890, 69074, 1306466, 25809826, 526358946, 10997782882, \dots
\end{flushleft}

\subsection{Unrestricted Noncrossing Digraphs}

To obtain a deduction system for \emph{unrestricted} (not necessarily
acyclic) noncrossing digraphs, one does away with the items for
minmax"-connected and maxmin"-connected graphs, and deletes all rules
that reference them -- with the exception of rules 01 and 02
(concatenating two edge"-covered graphs with the same directionality
of the covering edge), which should be changed to produce
mix"-connected items.
This yields the following integer sequence for unrestricted
noncrossing digraphs:
\begin{flushleft}
  1, 4, 64, 1792, 62464, 2437120, 101859328, 4459528192, 201889939456, \dots
\end{flushleft}

\subsection{Noncrossing Undirected Graphs}

By removing rules 24--26 (or 21--23), one obtains a tabulation of
noncrossing \emph{undirected} graphs.
This is also known as the class of (undirected) graphs with
pagenumber~1 under a fixed ordering of the vertices along the spine.
This class is counted by A054726 resp.\ A007297 (if additionally one
requires the graph to be connected) in the OEIS \citep{oeis}:
\begin{flushleft}
  1, 1, 2, 8, 48, 352, 2880, 25216, 231168, 2190848, \dots\\
  1, 4, 23, 156, 1162, 9192, 75819, 644908, 5616182, 49826712, \dots
\end{flushleft}

\section*{Acknowledgments}

The decomposition that is the basis for the tabulation presented in
this note was inspired by a counting technique for noncrossing acyclic
digraphs proposed by \citet{tirrell2014number}.
I benefited greatly from discussions with the participants of the
Dagstuhl Seminar 15122 `Formal Models of Graph Transformation in
Natural Language Processing'.

\section*{Document history}

\begin{description}
\item[2014-08-12] First version.
\item[2014-08-13] Add a section on derived tabulations (weakly
  connected noncrossing digraphs, unrestricted noncrossing digraphs).
\item[2015-04-20] Add a mention that the tabulation can be used to
  count the number of noncrossing undirected graphs.
\end{description}

\bibliographystyle{plainnat}
\bibliography{mcqm}

\end{document}